\pdfoutput=1
\documentclass[11pt,twocolumn]{article}

\usepackage{times}
\usepackage[
    left=1.5cm,
    right=1.5cm,
    top=3.00cm,
    bottom=3.00cm
]{geometry}

\usepackage[normalem]{ulem}
\usepackage[abbreviations]{foreign}  % For \ie, \eg \etc
\usepackage{fdsymbol}
\usepackage{booktabs}
\usepackage[T1]{fontenc}
\usepackage{pifont}
\usepackage{listings}
\usepackage{xspace}
\usepackage{ifthen}
\usepackage{setspace}
\usepackage[inline]{enumitem}
\usepackage{acronym}
\usepackage{framed}
\usepackage[usenames, dvipsnames]{xcolor}
\usepackage{framed}
\definecolor{shadecolor}{RGB}{240,240,240}
\usepackage{graphicx} % to include graphics
\usepackage{caption}

\newcommand{\myparagraph}[1]{\vspace{1mm} \smallskip \noindent{\bf {#1}}}

\usepackage{etoolbox}
\usepackage{url}

\usepackage{breakurl}

\usepackage[]{hyperref}
\hypersetup{
  colorlinks,
  linkcolor={green!50!black},
  citecolor={red!70!black},
  urlcolor={blue!70!black}
}

\usepackage[tracking=true]{microtype}
\newcommand{\sys}{\textls[-50]{\textsc{Wawel}}\xspace}

\acrodef{VEE}{virtual execution environment}
\acrodef{VSE}{virtual secure element}
\acrodef{IMA}{Linux integrity measurement architecture}
\acrodef{TPM}{trusted platform module}
\acrodef{HSM}{hardware security module}
\acrodef{PCR}{platform configuration register}
\acrodef{CA}{certificate authority}
\acrodef{SEV}{Secure Encrypted Virtualization}
\acrodef{gRPC}{remote procedure calls}
\acrodef{TLS}{transport layer security}
\acrodef{mTLS}{mutual transport layer security}
\acrodef{SE}{Secure Execution}
\acrodef{AWS}{Amazon Web Services}
\acrodef{TSS}{TPM software stack}
\acrodef{TDX}{Trust Domain eXtensions}
\acrodef{SGX}{Software Guard eXtensions}
\acrodef{TEE}{trusted execution environment}
\acrodef{RSA}{Rivest-Shamir-Adleman}
\acrodef{FIPS}{federal information processing standard}
\acrodef{DICE}{device identifier composition engine}
\acrodef{CRTM}{core root of trust for measurement}
\acrodef{DCAP}{data center attestation primitives}
\acrodef{LPAR}{logical partitions}
\acrodef{MAC}{message authentication code}
\acrodef{PEF}{Protected Execution Facility}
\acrodef{VTPM}[vTPM]{virtual TPM}
\acrodef{TCB}{trusted computing base}
\acrodef{API}{application programming interface}
\acrodef{UEFI}{unified extensible firmware interface}
\acrodef{RoT}{root of trust}
\acrodef{HMAC}{hash-based message authentication code}
\acrodef{EP11}{Enterprise PKCS\#11}
\acrodef{KVM}{kernel-based virtual machine}
\acrodef{PKCS11}[PKCS\#11]{public key cryptography standard \#11}
\acrodef{TCG}{Trusted Computing Group}
\acrodef{IoT}{internet of things}
\acrodef{IP}{intellectual property}
\acrodef{EK}{endorsement key}
\acrodef{UDS}{unique device secret}

\begin{document}

\title{Scalable Attestation of Virtualized Execution Environments \\in Hybrid- and Multi-Cloud}

\author{
        Wojciech Ozga\\
        \textit{IBM Research Zurich}
     \and
        Patricia Sagmeister\\
        \textit{IBM Research Zurich}
     \and
        Tamás Visegrády*\thanks{*Tamás Visegrády is now affiliated with Metaco Labs}\\
        \textit{IBM Research Zurich}
    \and
        Silvio Dragone\\
        \textit{IBM Research Zurich}        
}

\maketitle

\begin{abstract}
Existing attestation mechanisms lack scalability and support for heterogeneous \acp{VEE}, such as virtual machines and containers executed inside or outside hardware isolation on different vendors' hardware in clouds managed by various organizations. To overcome these limitations, hardware vendors and cloud providers implement proprietary mechanisms (Intel DCAP, Amazon NitroTPM, Google Titan) to support their offerings. However, due to their plurality, the attestation becomes cumbersome because it increases maintenance and integration costs and reduces portability required in hybrid- and multi-cloud deployments.

We introduce \sys, a framework that enables scalable attestation of heterogeneous \acp{VEE}. \sys can be plugged into existing hardware-specific attestation mechanisms, offering a unified interface. \sys supports the widely adopted \ac{TPM} attestation standard. We implemented a prototype and integrated it with three different \acp{VEE}. It supports runtime integrity attestation with \ac{IMA} and legacy applications requiring zero-code changes. The evaluation demonstrated that the \sys prototype achieves very good performance and scalability despite the indirections between the \ac{VEE} and hardware root of trust.

\end{abstract}
\acresetall
\section{Introduction}

Cloud providers and tenants attest to the computing environment to ensure its compliance with security requirements. They use attestation protocols to obtain technical assurance that computations execute on the required operating system running on specific hardware in the cloud. Unfortunately, the existing attestation mechanisms do not scale \cite{tcg2016tpm, intel2016epid}, are limited to particular vendors' hardware \cite{savagaonkar2017titan, intel2018dcap, intel2016epid, hamilton2021nitro}, or do not support different virtualization technologies \cite{perez2006vtpm, savagaonkar2017titan}.
We address these problems with an architecture that offers scalability and portability, properties that are required in hybrid- and multi-cloud deployments. Our approach unifies vendor- and virtualization-specific mechanisms behind an open and standardized attestation interface, leading to decreased maintenance and integration costs.

The virtualization technology enables the economic value of clouds because physical resources are dynamically provisioned and shared among tenants accordingly to their business needs. Logically, a single powerful physical computer creates smaller logical computational units called \acp{VEE}. Based on the non-functional requirements, such as performance, resource allocation, and isolation level, clouds create \acp{VEE} using different virtualization and isolation techniques. These offer different security guarantees. Our architecture, \sys, operates under the threat model of the virtualization technology underpinning the \ac{VEE}. Examples of \acp{VEE} are containers \cite{soltesz2007containers}, virtual machines \cite{bellard2005qemu, agache2020firecracker}, isolated processes (enclaves) \cite{costan2016sanctum, costan2016intel}, isolated containers \cite{arnautov2016scone}, or secure virtual machines \cite{tdxwhitepaper, kaplan2017sev}.

Modern computing software stacks, like \ac{UEFI} \cite{wilkins2013uefi, tcg_srtm_bios_spec, tcg_srtm_uefi_spec}, Linux kernel \cite{linux}, and Microsoft Windows, adopted the \ac{TPM} standard \cite{tcg2016tpm} that laid the foundations for the integrity attestation of personal computers \cite{tcg_tpm_attestation}. Today's firmware and software use \ac{TPM}-compatible devices to securely store and certify integrity measurements of the boot firmware and operating system. Although widely adopted, this technology failed in cloud deployments due to scalability limitations. Specifically, the demand for the secure storage capacity for storing integrity measurements of \acp{VEE} grows faster than the physical storage capacity of a resource-constrained discrete \ac{TPM} device. 
 
To overcome this limitation, cloud providers turn to proprietary hardware to enable attestation at scale \cite{hamilton2021nitro, savagaonkar2017titan, intel2018dcap, lagar2020caliptra}. However, proprietary hardware increases infrastructure costs and forces tenants to adapt to the cloud vendor's \ac{API}. \sys overcomes these limitations. It relies on a limited number of cryptographic coprocessors that handle security-sensitive attestation operations in a high-availability and fault-tolerant way while relying on the state-of-the-art \ac{TPM} protocol's \ac{API} for portability. Unlike \ac{TPM} devices that are hard linked with a physical computer, \sys introduces stateless cryptographic coprocessors that offer \ac{TPM}-like capabilities to a virtually unlimited number of \acp{VEE}. The scalability results from the secure offloading of the \ac{VEE}-specific state from the cryptographic coprocessor. Conversely, in the case of the \ac{TPM}, the state never leaves the \ac{TPM} device's secure boundary.

\sys provides attestation primitives that can be plugged into different virtualization and isolation technologies. The advantage of \sys is that it complements existing vendor-specific attestation mechanisms with the support of an open attestation standard. Consequently, tenants can use the standardized protocol \cite{tcg_tpm_attestation} and its large software and hardware ecosystem \cite{ibm_tpm_tss, sailer2004ima, wilkins2013uefi, broz2018luks, strongswan_org, tcg_srtm_bios_spec, tcg_srtm_uefi_spec} to collect technical assurance of the provisioned \acp{VEE} regardless of the virtualization type (virtual machine \cite{bellard2005qemu}, container \cite{soltesz2007containers}), isolation level (confidential computing \cite{hunt2021confidential, tdxwhitepaper, kaplan2017sev, costan2016intel, borntraeger2020se}), or cloud-specific root of trust (Titan \cite{savagaonkar2017titan}, NitroTPM \cite{hamilton2021nitro}).

\sys has noteworthy advantages. First, unlike state-of-the-art \ac{TPM} devices, \sys provides scalability because it introduces the concept of stateless cryptographic coprocessors that provide architectural primitives for attestation. Second, it enables portability by implementing the widely adopted \ac{TPM} protocol's \ac{API}, abstracting tenants from the underlying hardware and cloud provider-specific attestation. Third, it suits hybrid- and multi-cloud deployments by offering a unified attestation interface based on the distributed root of trust.

Evaluation of the \sys prototype showed that it is practical in terms of portability, performance, and scalability. It provides a drop-in replacement for \ac{TPM}-based architectures, supporting \ac{TPM}-based applications with zero-code changes. We integrated the \sys prototype with three different \acp{VEE}. Furthermore, the prototype supports runtime integrity measurement and attestation with the help of \ac{IMA} and a custom TPM driver. The \sys prototype achieved low latencies for retrieving the signed attestation quote, 5\,ms vs 209\,ms, and extending integrity measurements, 1\,ms vs 9\,ms, compared to a hardware \ac{TPM}, respectively. A prototype equipped in a single stateless cryptographic coprocessor achieved a throughput of 20k integrity measurement extensions per second and 7.5k quotes per second that justifies the economical use of the proposed design in clouds.

In summary, we make the following contributions:
\begin{itemize}[noitemsep, leftmargin=4mm]
    \item We introduced a novel approach to scalable attestation by introducing the offloading of integrity measurements from stateless cryptographic coprocessors (\S\ref{sec:overview}).
    \item We analyzed and discussed the security rationale related to the offloaded integrity measurements (\S\ref{sec:security_rationale}). 
    \item We designed \sys, a scalable attestation architecture of heterogenous \acp{VEE} that suits hybrid- and multi-cloud requirements (\S\ref{sec:overview}).
    \item We implemented a prototype of the \sys architecture that supports three different \acp{VEE}, as well as \ac{TPM}-based applications and Linux \ac{IMA} (\S\ref{sec:implementation}).
    \item We evaluated the performance and scalability of the \sys prototype (\S\ref{sec:evaluation}).
\end{itemize}

\acresetall
\section{Overview}
\label{sec:overview}

\subsection{High-level Overview}
\begin{figure}[tbp!]
    \centering
    % \captionsetup{skip=4pt}
    \includegraphics[width=0.48\textwidth]{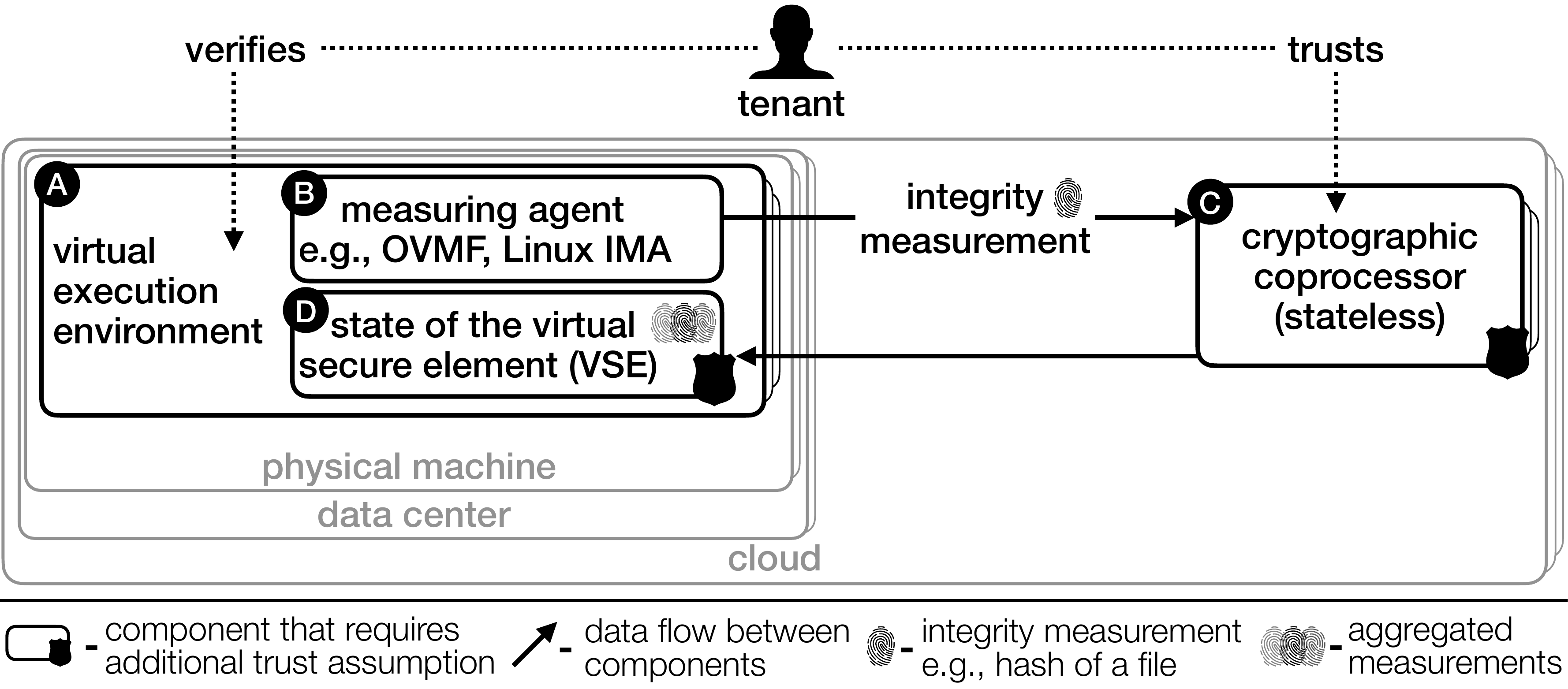}
    \caption{
        High-level overview of the \sys design. To achieve scalability, stateless cryptographic coprocessors securely offload the \acf{VSE} state to \acf{VEE}. A \ac{VSE} state contains attestation-related \ac{VEE}-specific information, like integrity measurements.
    }
    \label{fig:overview}
\end{figure}

\autoref{fig:overview} shows the high-level overview of the \sys architecture that consists of four components: 
\begin{description}
    \item[(A)] \Acp{VEE} that are subject of attestation. For example, virtual machines or containers.
    \item[(B)] A measuring agent, a piece of code that collects and records the integrity measurements.
    \item[(C)] Stateless cryptographic coprocessors that act as a \acf{RoT} for attestation, for example, network-accessible \acp{HSM}.
    \item[(D)] \Acf{VSE} states, cryptographically protected states containing \acp{VEE}' integrity measurements.
\end{description}

A tenant trusts cryptographic coprocessors; based on this relation, he establishes trust with the \ac{VEE}. First, he retrieves a certificate containing the \ac{VEE}'s integrity measurements. Then, he ensures that the genuine cryptographic coprocessor signed this certificate and the certified measurements reflect the expected \ac{VEE}.

Cryptographic coprocessors enable attestation of \acp{VEE} by providing the necessary architectural primitives. These are secure storage aggregating integrity measurements and a signing mechanism that certifies these measurements. Each VEE (A) has its own VSE instance, which consists of the VSE state (D) and a stateless cryptographic coprocessor (C). For scalability purposes, unlike in the existing attestation designs \cite{perez2006vtpm, savagaonkar2017titan}, the VSE state is securely offloaded from the cryptographic coprocessor. Thus, stateless cryptographic coprocessors can handle a virtually unlimited number of VEEs. We discuss the security rationale of the offloaded \ac{VSE} state in \S\ref{sec:security_rationale}.

Cryptographic coprocessors are high-assurance devices optimized for high-throughput, low-latency cryptographic operations. Examples of cryptographic coprocessors are \acp{HSM} that build the security foundation of modern clouds and serve as the \ac{RoT} for safety- and security-critical applications, like banking or governmental systems. \acp{HSM} are certified at \ac{FIPS} 140-2 Level 3 or Level 4 \cite{fips140_2} that respond to unauthorized access at software and physical level. Therefore, \sys uses them as a \ac{RoT} component, extending them with the \ac{VSE} functionality that enables scalable \acp{VEE} attestation.

\begin{figure}[tbp!]
    \centering
    % \captionsetup{skip=4pt}
    \includegraphics[width=0.48\textwidth]{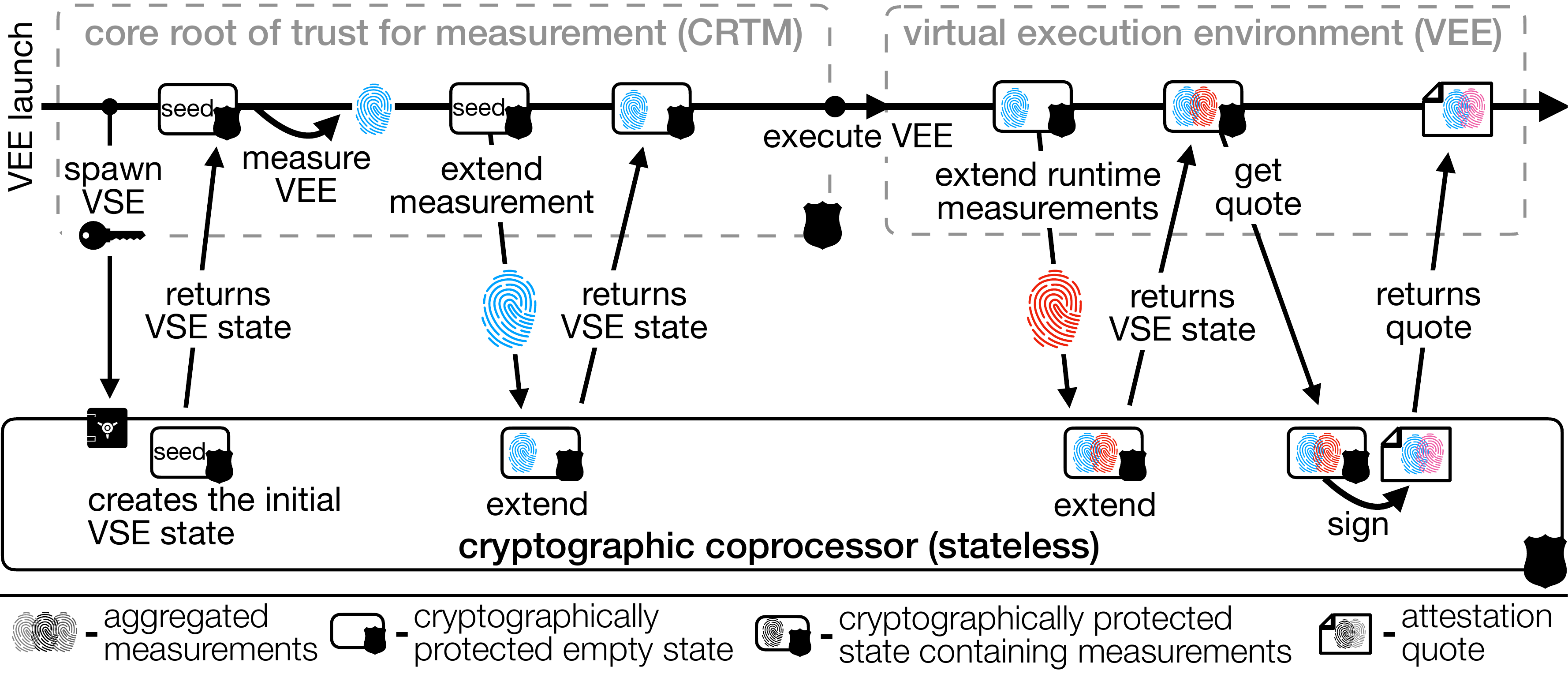}
    \caption{
        Integrity measurement and quote generation process. Measurement agents collect integrity measurements and record them in the \acf{VSE}. Agents execute sequentially. The first one, called \acf{CRTM} initializes the chain of trust and, therefore, is trusted.
    }
    \label{fig:trustedboot}
\end{figure}

\subsection{Integrity Measurements}
Each \ac{VEE} logically possesses a dedicated \ac{VSE} instance, which stores integrity measurements and certifies them with a digital signature. Technically, a cryptographic coprocessor creates a \ac{VSE} on behalf of a trusted process called \acf{CRTM}. \autoref{fig:trustedboot} shows how CRTM starts the measurement chain. Together with subsequent processes (measurement agents), it collects integrity measurements and sends them to the cryptographic coprocessor to persist them in the \ac{VSE} state attached to the request. Measurement agents always replace the previous state with the latest one to prevent rollbacks to the previous \ac{VSE} state (see \S\ref{sec:relay_attack}), similarly to the systems implementing the TCG's attestation schemes (see \S\ref{sec:security:schemes}).

\subsection{Attestation}
The verifier, like a tenant owning a VEE or an auditor, attests to the \ac{VEE} integrity to ensure that the \ac{VEE} conforms with security requirements imposed by regulators or a security policy. He requests a proof from the cryptographic coprocessor in the form of a digitally signed certificate. This certificate includes a digest over aggregated integrity measurements, a nonce, and a signature issued with the cryptographic coprocessor signing key. The verifier checks that the genuine cryptographic coprocessor signed the certificate, the nonce guarantees the freshness, and the cryptographic hash over aggregated measurements corresponds to a \emph{golden} hash representing the trusted integrity state of the \ac{VEE}.
Each VSE has a unique random seed that acts as the VEE identity. This seed is embedded in the VSE-specific \ac{EK} certificate linked to the cryptographic coprocessor's manufacturer certificate. The verifier uses the seed to differentiate among VEEs, similarly how hardware TPM's EK are used.

\subsection{Unified Attestation Interface}
The \ac{TPM} specification \cite{tcg_tpm_attestation} is the open standard that defines the attestation protocol. The industry has widely adopted it across the entire firmware and software stack \cite{wilkins2013uefi, linux, sailer2004ima, broz2018luks, strongswan_org}. However, alternative attestation mechanisms emerged for clouds because of the scalability limitation of the \ac{TPM} architecture that requires each execution environment to have a single hardware \ac{TPM} device. Standard protocols for cryptographic coprocessors, such as \ac{PKCS11} \cite{oasis240pkcs11} and \ac{EP11} \cite{visegrady2017ep11}, do not implement the \ac{TPM} protocol. 

To maintain a unified attestation interface and to overcome the scalability limitations, \sys extends cryptographic coprocessors with two functionalities. The first one is the support for the minimal subset of the \ac{TPM} protocol required for attestation: the generation of a signed quote and secure management of integrity measurements. The second functionality is the offloading of the state containing cryptographic measurements to enable scalability by maintaining stateless security coprocessors.

\section{Security Rationale}
\label{sec:security_rationale}

\subsection{Attestation schemes}
\label{sec:security:schemes}
\begin{figure}[tbp!]
    \centering
    % \captionsetup{skip=4pt}
    \includegraphics[width=0.48\textwidth]{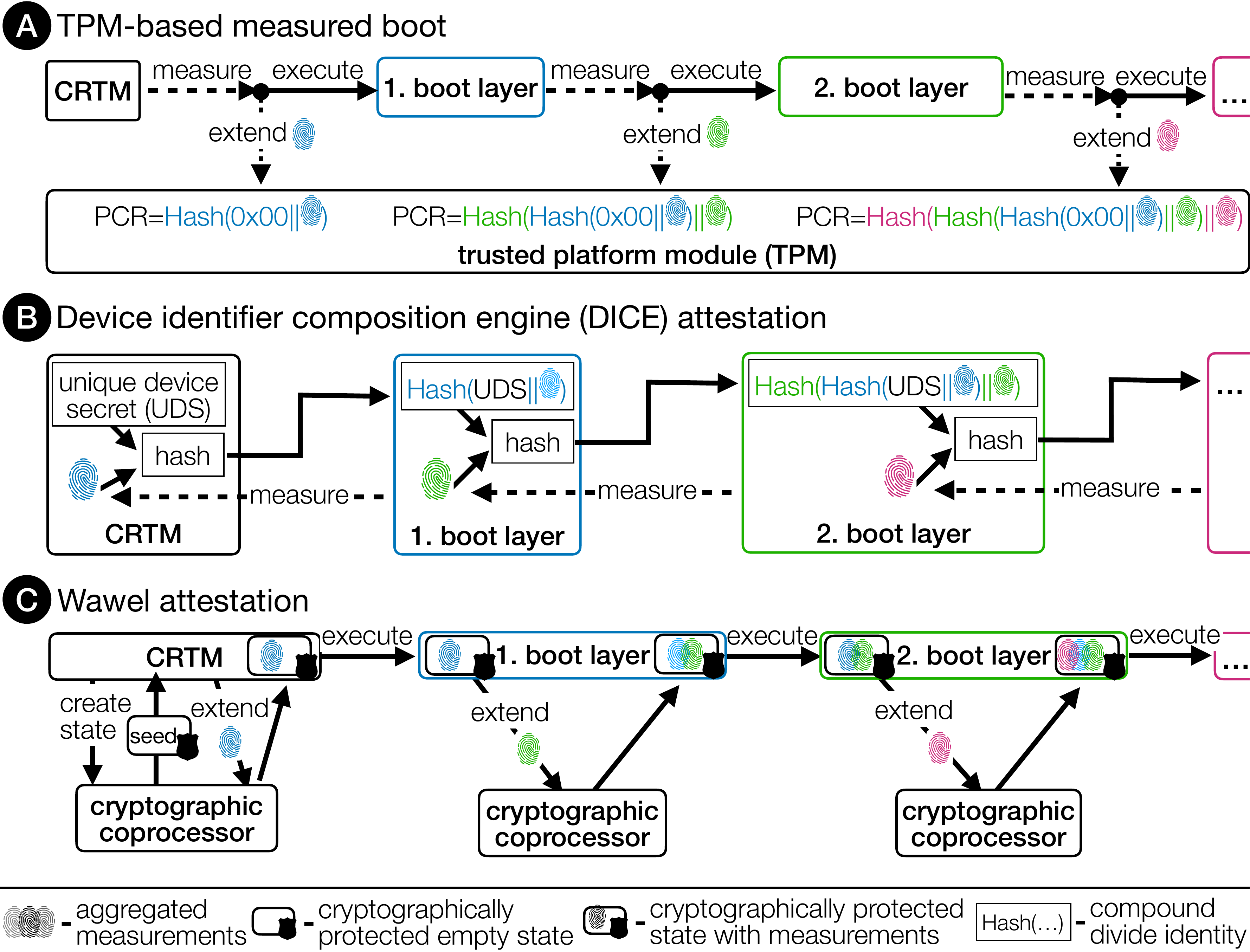}
    \caption{
        Comparison of the \sys attestation scheme with the TCG measured boot and DICE. \sys benefits from both designs. It relies on an external secure element for security-critical operations while the state is being kept outside the secure element within the measuring agent.
    }
    \label{fig:attestations}
\end{figure}

The \sys architecture leverages concepts from the state-of-the-art \ac{TCG}'s attestation schemes, namely the measured boot~\cite{tcg_tpm_attestation} and the \ac{DICE}~\cite{tcg2020dice}. Following the measured boot approach, \sys requires access to a secure element offering security-relevant features, such as the digital signing and cryptographic aggregation of integrity measurements. Unlike the TPM-based approach, \sys requires a stateless secure element for scalability reasons, moving \sys toward \ac{DICE}, where each measuring agent manages the aggregated measurement and moves it to the next measuring agent executing in the subsequent boot level.

\autoref{fig:attestations} presents the overview of TCG attestation schemes and \sys. They all rely on the chain of trust execution that starts with the trusted measuring agent called \ac{CRTM}. In the TCG's schemes, CRTM is baked into the read-only memory inside the boot processor in the measured boot and DICE designs. \sys instead requires implementing the CRTM inside the \ac{TCB} of the target virtualization technology (see \S\ref{sec:related_work}). 

The measured boot (A) relies on the \ac{TPM}~\cite{tcg2016tpm} to securely store and sign measurements. The TPM implements \acfp{PCR} that can only be cryptographically extended with a new measurement and reset only during the processor's power-on cycle. Measuring agents, starting with the CRTM, extend PCRs with measurements of the following boot layer before executing it. Therefore, PCRs reflect the integrity state of the executed software because an adversary cannot revert or overwrite the values of PCRs. The TPM stores the \ac{EK}, which is a cryptographic identity that uniquely identifies the TPM and signs the measurements stored in PCRs.

The DICE architecture (B) originally targeted systems that cannot use the TPM due to cost or size restrictions, like \ac{IoT}. However, DICE's simplicity brought it popularity, leading to the formally proven implementation~\cite{tao2021dice} and dedicated \ac{IP} block for system on chip (SoC)~\cite{lagar2020caliptra}. The key idea of the DICE design is that CRTM initializes the aggregated measurements with the \ac{UDS} and every subsequent measuring agent measures the next boot layer and aggregates this measurement using the one-way hash function~\cite{hash_functions}. The result is passed between the measuring agents of subsequent boot layers. The verifier can then establish trust with the software that proves posession of the aggregated measurement derived from the UDS. A more complex DICE variant involves asymmetric cryptography. Consecutive boot layers receive the private key and certificate containing the aggregated measurements from the previous layers. The verifier uses a certificate chain linked to the device manufacturer to verify the genuineness of the device and checks the aggregated measurement inside the leaf certificate.

The \sys architecture (C) derives from both concepts. The cryptographic coprocessor implements security-sensitive functionalities, like measurement aggregation using the hash function and digital signing of the quote. However, the state containing integrity measurements and a unique seed used to differentiate among VEEs is offloaded from the cryptographic coprocessor to measuring agents. Like in the DICE design, each measuring agent passes the state to the next boot layer and removes the previous version of the state before passing control to the next boot layer. 

\subsection{Offloaded state}
\begin{figure}[tbp!]
    \centering
    % \captionsetup{skip=4pt}
    \includegraphics[width=0.48\textwidth]{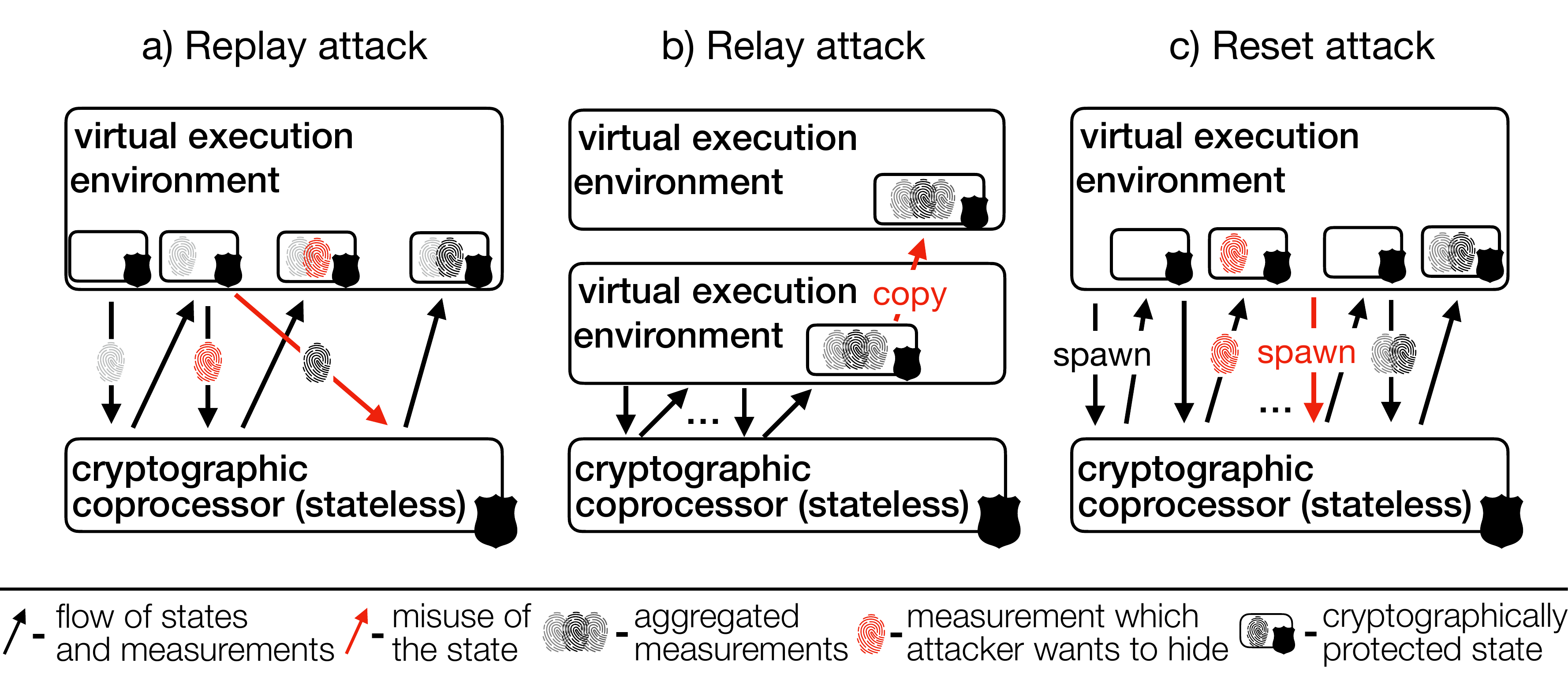}
    \caption{
        Potential misuse of the cryptographically protected state that leads to attacks discussed in \S\ref{sec:security_rationale}.
    }
    \label{fig:attacks}
\end{figure}

In the \sys architecture, cryptographic coprocessors create and manage \acp{VSE}. However, because of the limited physical storage, they offload the \ac{VEE}-specific state. \acp{VEE} attach the state to every request they make to cryptographic coprocessors. The offloaded state is protected against tampering with a \ac{MAC}. A cryptographic coprocessor attaches to the \ac{VSE} state a cryptographic signature --- an \ac{HMAC}, which it then verifies before using it~\cite{visegrady2014stateless}.

Although the \ac{VSE} state is protected against tampering, we analyze in this section potential attacks that an adversary could run when having access to legitimate, correctly signed \ac{VSE} states.  

\myparagraph{Shared HMAC key.}
Cryptographic coprocessors share the same \ac{HMAC} key for scalability, high availability, and fault-tolerance reasons. Sharing the same \ac{HMAC} key allows them to handle requests produced by related cryptographic coprocessors. The administrator increases and decreases the number of cryptographic coprocessors according to high availability and scalability requirements without affecting security. 

The \ac{HMAC} key never leaves cryptographic coprocessors. The administrator requests its creation and then distributes it to cryptographic coprocessors to form the pool of cryptographic coprocessors serving attestation primitives. The administrator never sees the \ac{HMAC} key in plain text. He uses one cryptographic coprocessor to generate the \ac{HMAC} key and then offloads it. The offloaded key is wrapped with a master key of another cryptographic coprocessor. The administrator imports then the wrapped key to another cryptographic coprocessor, which unwraps the \ac{HMAC} key using its master key.

\myparagraph{Replay attacks.}
\autoref{fig:attacks}a) shows the replay attack. In this attack, an adversary controlling a malicious \ac{VEE} replays an old \ac{VSE} state to create a deviating measurement chain that does not include all measurements. There are two different approaches to mitigate this attack. In the first one, the \ac{VEE} removes old \ac{VEE} states. In the second one, the cryptographic coprocessor leverages monotonic counters for freshness protection. 

The first approach consists of correctly implementing measurement agents inside the \ac{VEE}. They must destroy the old \ac{VSE} state immediately after they receive the new one and before they execute the measured code. Otherwise, the malicious code, which was measured, has access to the old \ac{VSE} state that does not include her measurement. She can vanish the measurement by reusing the old state, effectively hiding her presence from the verifier. It is the same assumption as in the \ac{DICE}~\cite{tcg2020dice}. 

The second approach leverages monotonic counters \cite{matetic2017rote, strackx2016ariadne} that version each \ac{VSE} state. The cryptographic coprocessor adds the value of the monotonic counter to the \ac{VSE} state and increments the counter. The next time it receives the \ac{VSE} state, it compares the monotonic counter value with the value stored in the \ac{VSE} state. A mismatch indicates the replay attack. 

\myparagraph{Relay attacks.}
\label{sec:relay_attack}
\autoref{fig:attacks}b) illustrates the relay attack. The adversary who gained temporal control over the \ac{VEE} copies the \ac{VSE} state to another \ac{VEE}. He can then certify the other \ac{VEE}'s integrity using the \ac{VSE} state, even though this state does not reflect the true integrity of the other \ac{VEE}. This attack is equivalent to relay attacks, like the TPM cuckoo attack \cite{parno2008cuckoo}, in which an adversary responds to attestation queries using the \ac{TPM} of a legitimate \ac{VEE}. Please note that the certified integrity measurements in this attack still reflect the \ac{VEE} integrity. The \ac{VEE} owner is responsible for protecting the \ac{VEE} and its state and not exposing the attestation functionality to the untrusted outside world.

\myparagraph{Reset attacks.}
\label{sec:reset_attack}
An adversary might want to impersonate a legitimate \ac{VEE} by convincing a tenant that the \ac{VEE} under her control is legitimate. To achieve it, she could exploit the attestation infrastructure to obtain proof containing \emph{golden measurements} that correspond to the legitimate \ac{VEE}. 

\autoref{fig:attacks}c) depicts the attack in which an attacker requests a new \ac{VSE} that does not contain any measurement yet. Then, simulating the behaviour of legitimate measurement agents, she replays the golden measurements. This attack is equivalent to the \ac{TPM} reset attack \cite{reset2007sparks, kauer2007oslo}, in which an adversary gets rid of unwanted measurements by first resetting the TPM device and then sending arbitrary measurements.

Two alternative solutions exist to mitigate this attack. In the first one, the cryptographic coprocessor initializes the \ac{VSE} state with a random number measurement. This mitigates the attack because the probability that an adversary finds measurements leading to the desired digest included in the attestation is negligible because of the soundness of the cryptographic hash-function properties \cite{hash_functions, dif77}. In this approach, however, the random number must be known to the tenant, so he can recalculate the attestation digest. One option is that the cloud provider discloses the random number during the \ac{VEE} provisioning.

In the second approach, authorizing access to the cryptographic coprocessors' interface prevents an adversary from creating \acp{VSE}. \Acp{CRTM} are then the only components allowed to create new \acp{VSE}. They identify themselves using credentials, like \ac{TLS} credentials or pre-shared \ac{TLS} keys. To differentiate among execution environments with different security assurance, \acp{CRTM} possess different credentials. Cryptographic coprocessors sign attestation proofs using keys to differentiate among virtualization technologies.
\section{Implementation}
\label{sec:implementation}
\begin{figure}[tbp!]
    \centering
    % \captionsetup{skip=4pt}
    \includegraphics[width=0.48\textwidth]{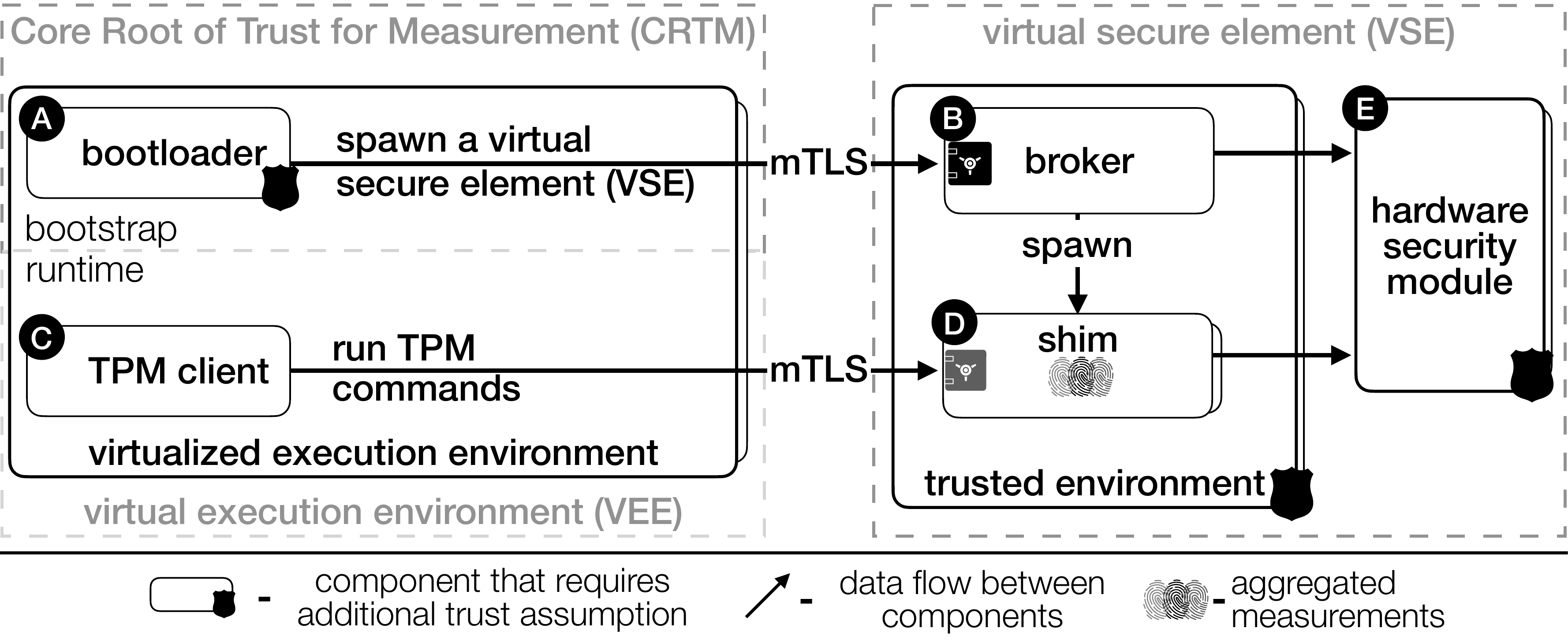}
    \caption{
        High-level overview of the \sys prototype. 
    }
    \label{fig:implementation}
\end{figure}

\autoref{fig:implementation} shows the architecture of the \sys prototype implementation, which is compatible with the \ac{TPM} protocol \cite{tcg2016tpm} and supports the  \ac{IMA} \cite{sailer2004ima}. The prototype consists of five components:
\begin{description}
    \item[(A)] Bootloader firmware implementing CRTM and requesting a new \ac{VSE},
    \item[(B)] Broker service creating new \acp{VSE},
    \item[(C)] Client executing inside of a \ac{VEE} and using the \ac{VSE} for attestation,
    \item[(D)] Shim translating the TPM protocol to the \ac{HSM} protocol,
    \item[(E)] \acp{HSM} supporting the \ac{VSE} extension.
\end{description}

The \sys prototype implements the shim because we could not modify the \ac{HSM} firmware to fully implement the required architectural changes. The shim is a layer of indirection between the client and the \ac{HSM} that performs security-critical operations that should be otherwise implemented in the HSM. Specifically, it constructs the TPM-compatible quote structure containing expected integrity measurements. It then requests the HSM to sign the quote.
 
We implemented the majority of the \sys components in the memory-safe language Rust~\cite{matsakis2014rust}. We used the C language to implement the custom Linux kernel driver. We relied on the \emph{m\_SetAttributeValue} of the \ac{EP11} \cite{visegrady2017ep11} command set to emulate the \ac{PCR} functionality in the \ac{HSM}. The prototype was integrated with three \acp{VEE}: Docker containers, QEMU \ac{KVM} \cite{bellard2005qemu}, and one hardware-based \acf{TEE} technology.\footnote{Details omitted due to double-blindness requirements}

\subsection{Load-time integrity measurements}
\begin{figure}[tbp!]
    \centering
    % \captionsetup{skip=4pt}
    \includegraphics[width=0.48\textwidth]{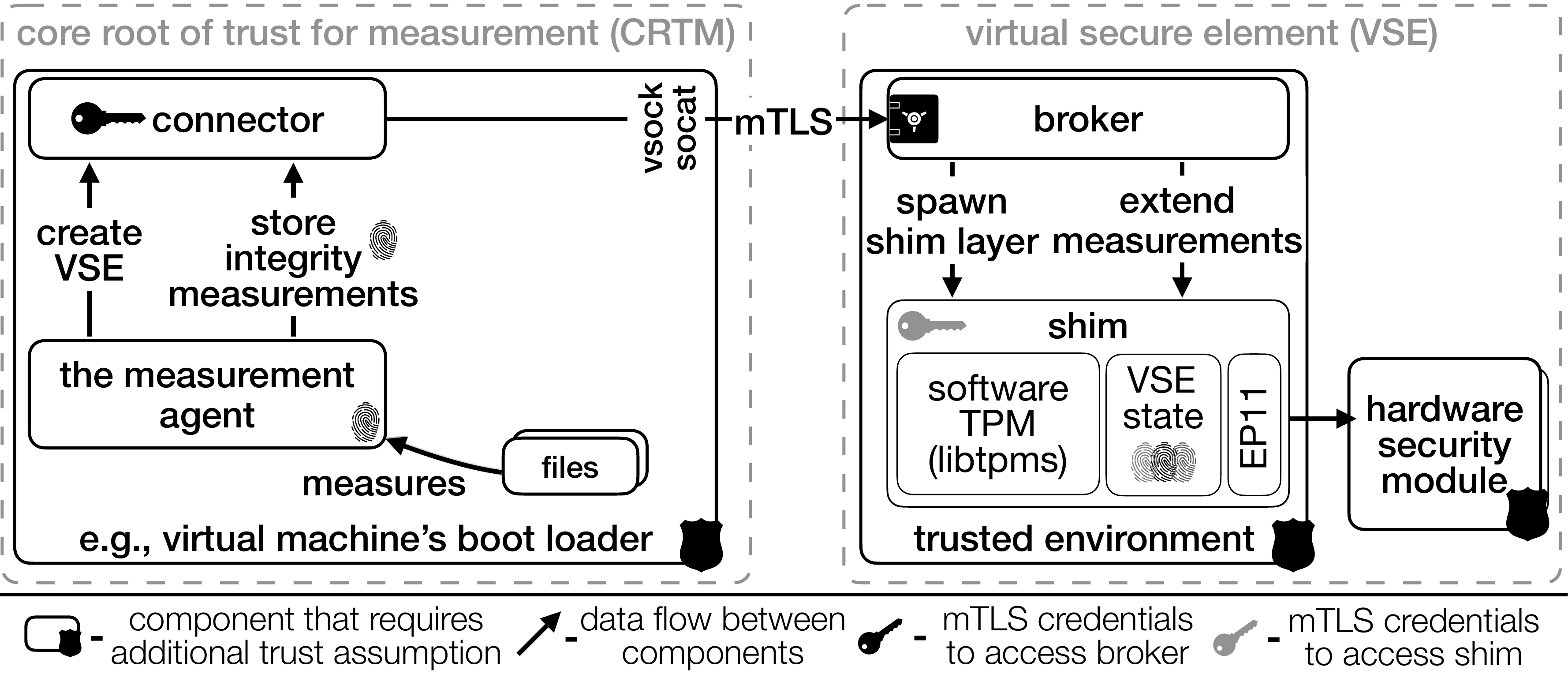}
    \caption{
        \sys prototype implementation of the \acf{VEE} bootstrap. The \acf{CRTM} creates a new \acf{VSE} and populates it with early integrity measurements before passing control to untrusted runtime components. 
    }
    \label{fig:bootstrap}
\end{figure}

\autoref{fig:bootstrap} shows the creation of a \ac{VSE} instance and the extension of first integrity measurements during the boot process of the \ac{VEE}. These security-critical operations must be performed within a trusted environment. Otherwise, an adversary could spawn a new \ac{VSE} to achieve an equivalent of the \ac{TPM} reset attack \cite{reset2007sparks, kauer2007oslo}, see \S\ref{sec:relay_attack}. Depending on the final environment, this could be implemented in, for example, a more privileged layer, like \emph{VMPL0} in the AMD \ac{SEV} guest OS, the encrypted bootloader of IBM Z Secure Execution~\cite{borntraeger2020se}, or trusted hypervisor like in the vTPM design~\cite{perez2006vtpm}. See \S\ref{sec:related_work} for more details.

Creating a new \ac{VSE} instance requires access to the broker service before the startup of the network stack, \ie, before the \ac{VEE} becomes accessible to the hostile outside world. In our prototype, the connector utility establishes a mutual \ac{TLS} over \ac{gRPC} to the broker service via the virtio-vsock interface \cite{hajnoczi2015vsock}. The vsock interface is available early during the boot process before the network is set up. Packages travel over the vsock (virtualized i/o) to the hypervisor and from there to the broker service running on the well-known host and port. 

Access to the broker service is mutually authenticated using \ac{TLS} credentials. Their distribution depends on the data center infrastructure. A \ac{CA} can, for example, distribute credentials statically during the \ac{VEE} image build, like in the case of the IBM Z Secure Execution technology \cite{borntraeger2020se}. Trusted hypervisors, like in the \ac{AWS} Nitro architecture \cite{hamilton2021nitro}, could unseal them from the hardware root of trust, \eg, a \ac{TPM}. \acs{TEE}, like AMD \ac{SEV} \cite{kaplan2017sev} or Intel \ac{TDX} \cite{tdxwhitepaper}, could recover them after the successful attestation of the initial bootloader.

\subsection{Runtime Integrity Measurements}
\begin{figure}[tbp!]
    \centering
    % \captionsetup{skip=4pt}
    \includegraphics[width=0.48\textwidth]{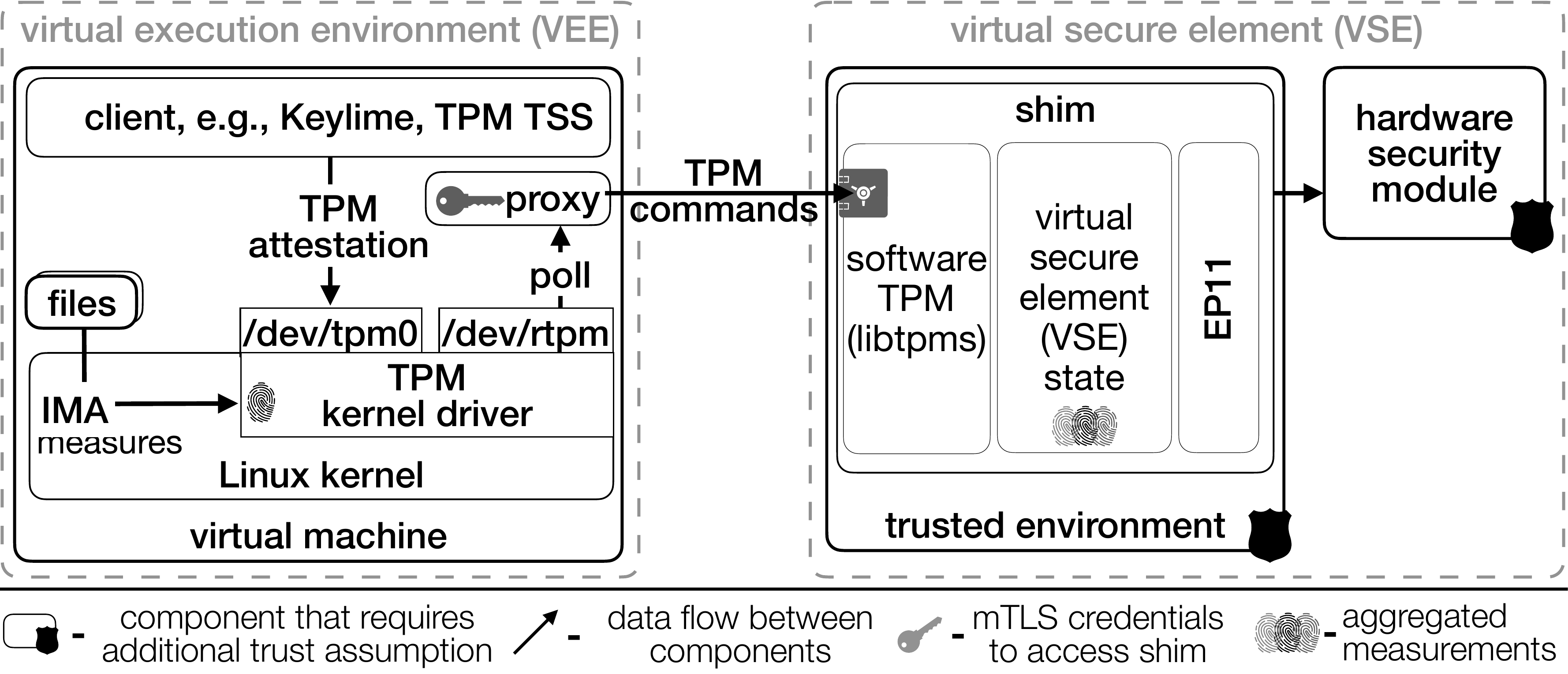}
    \caption{
        \sys prototype implementation with runtime components. The custom kernel driver and proxy hide the remote nature of the \acf{VSE}. Consequently, \sys transparently supports \ac{TPM} clients, like \acf{IMA}.
    }
    \label{fig:runtime}
\end{figure}

\autoref{fig:runtime} shows the use of the prototype \sys implementation inside of the \ac{VEE}. To support legacy TPM ecosystems, \sys implements a custom kernel driver. Consequently, TPM clients, such as the \ac{TSS} \cite{ibm_tpm_tss}, Keylime \cite{schear2016keylime}, and \ac{IMA} \cite{sailer2004ima}, access the \ac{VSE} as if they were accessing a locally attached hardware \ac{TPM}. They send TPM commands to the standard TPM interface controlled by a dedicated TPM driver (\emph{/dev/tpm0} device). A userspace application, called proxy, reads the TPM command from the kernel space using a dedicated interface (\emph{/dev/rtpm} device). The proxy component is required because the kernel driver running in the kernel space cannot establish the \ac{gRPC} network connection.

The TPM driver executes in two modes, asynchronous and synchronous mode. It starts in the asynchronous mode, aggregating TPM commands in the internal kernel buffer while responding with valid but made-up TPM responses. This allows the system to boot correctly despite lack of the link to the real TPM device. The driver switches to the synchronous mode as soon as the proxy utility sends all buffered TPM commands to the HSM. This happens early during the boot process when the HSM becomes reachable over the network. In the synchronous mode, the driver does not buffer commands anymore. It also returns the real TPM responses received from the shim.

The initial TPM commands to which the driver responds with made-up responses are sent by the Linux kernel to discover the TPM properties, like the number of \acp{PCR}, banks, or supported cryptographic algorithms. \ac{IMA} also sends measurements early during the system's boot and before the proxy is started. These initial \ac{IMA} measurements are also buffered and executed in the same order in which the proxy received them. Depending on the implementation, this might be a vulnerability window for the replay attack. One might eliminate or reduce the size of the vulnerability window by starting the proxy before the network setup. Like this, buffered measurements are processed before the adversary gets access to the \ac{VEE} over the network. The implementation might leverage, for example, the vsock interface to get access to the shim early in the boot process.
\section{Evaluation}
\label{sec:evaluation}

We evaluated \sys to answer the following questions:
\begin{enumerate}
    \item Is \sys practical in terms of the execution latency of TPM commands, compared to a hardware TPM?
    \item How many \acp{VEE} can be handled by a single stateless cryptographic coprocessor?
    \item Does the \sys architecture scale with the increasing number of \acp{VEE}?
\end{enumerate}

\textbf{Testbed:} Experiments execute on three machines located in the same data center and connected via 10\,Gb Ethernet. Machine A hosts QEMU \cite{bellard2005qemu} virtual machines and the \sys's broker service. Machine B runs \sys's shim services that have access to the \ac{HSM}. Machine C generates load during the scalability experiment.

\myparagraph{Machine A} is a Supermicro SYS-610C-TR server equipped with an Intel Xeon Gold 6338 CPU with 128 cores, 512 GiB of RAM, Infineon\,SLB-9670 TPM. This machine runs Ubuntu 22.04 with Linux kernel 6.1.0. 

\myparagraph{Machine B} is Trenton Systems 4U chassis with a HDB8231 backplane and a SEP8253 processor board with two Intel Xeon Silver 4109T @ 2\,GHz CPUs each one with 16 logical cores, 16\,GiB of RAM, and an IBM CryptoExpress7S. This machine runs CentOS Stream 8 with the Linux kernel 4.18.0.

\myparagraph{Machine C} is equipped with an Intel Xeon CPU E3-1275\,v3 with 8 cores, 32\,GiB of RAM. CentOS Stream 8 with Linux kernel 4.18.0 run on this machine. 

\subsection{Micro-benchmarks}

\begin{figure}[t!]
    \centering
    \includegraphics[width=0.5\textwidth]{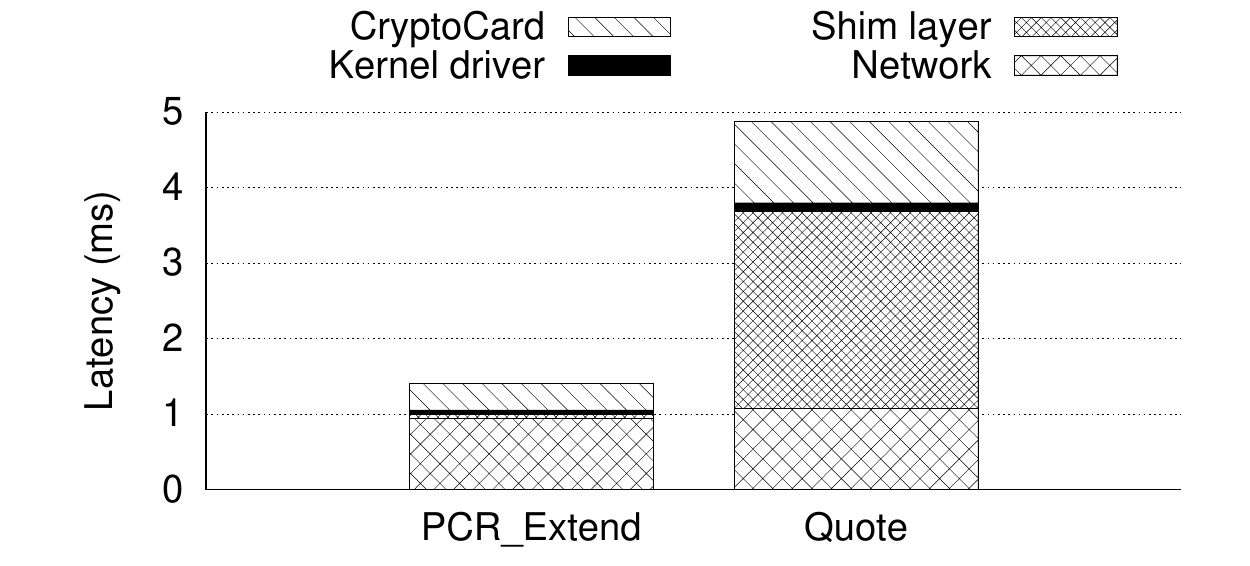}
    \caption{Composed latencies of running TPM commands on the \sys prototype.}
    \label{fig:eval:overheads}
\end{figure}
\myparagraph{What overhead is introduced by different components of \sys?}
We start the evaluation by profiling the \sys prototype to learn how much time is spend in different \sys components during the execution of two most frequently used TPM attestation-related commands: PCR Extend and Quote. To achieve that, we instrumented \sys prototype to measure the time taken by: \emph{Kernel driver}, \emph{Network}, \emph{Shim}, and \emph{CryptoCard}. The \emph{Kernel driver} represents the total time the TPM TSS tool takes to marshal and send the TPM command to the kernel driver, including the time the proxy takes to retrieve the command from the kernel space and process it. The \emph{Network} represents the total round-trip time taken for the command to travel between the proxy and the shim. The \emph{Shim} represents the execution time of the shim process, which includes parsing of the TPM command, translating between TPM and EP11 protocols, and generating the TPM response structure. The \emph{CryptoCard} represents the time taken by the IBM CryptoCard to process the request. For each TPM command, we ran the experiment ten times and used a 10\% trimmed mean to calculate the average execution latency.

\autoref{fig:eval:overheads} shows the latency of \sys components during the execution of the PCR Extend and Quote commands. The PCR Extend command, which is used to aggregate integrity measurements, takes 1.4\,ms. The Quote command takes more time 4.8\,ms because the shim must create first the complex TPM Quote structure and then the CryptoCard must sign it. The CryptoCard execution is one of the dominant factor in the total execution time. During the PCR Extend operation, the CryptoCard verifies the offloaded state using HMAC and then performs a single hash function, taking 0.3\,ms, or 25\% of the total command execution time. During the Quote command operation, the CryptoCard first verifies the offloaded state using HMAC and then performs an RSA-2048 signing, taking 1\,ms, or 22\% of the total command execution time. 

The second important component is the shim that implements the TPM protocol structures, taking 0.05\,ms and 2.6\,ms for PCR Extend and Quote, which corresponds to 3\% and 54\%, respectively. Network takes roughly a constant time of 1\,ms, which becomes 67\% and 22\% of the total execution time of these commands, respectively. The kernel driver, which enables support for legacy TPM applications via the standardized communication interface, takes 0.06\,ms and 0.11\,ms for PCR Extend and Quote, respectively. These correspond to 4\% and 2\% of the total command execution time, respectively. The low overhead of the kernel driver and proxy components is a result of the implementation design choices. The kernel driver implements the poll mechanism to inform the userspace proxy component about the newly arrived TPM command. The proxy component uses gRPC over HTTP2 that maintains a connection to the shim, reducing the need for TLS handshakes when sending TPM commands.

\begin{figure}[t!]
    \centering
    \includegraphics[width=0.5\textwidth]{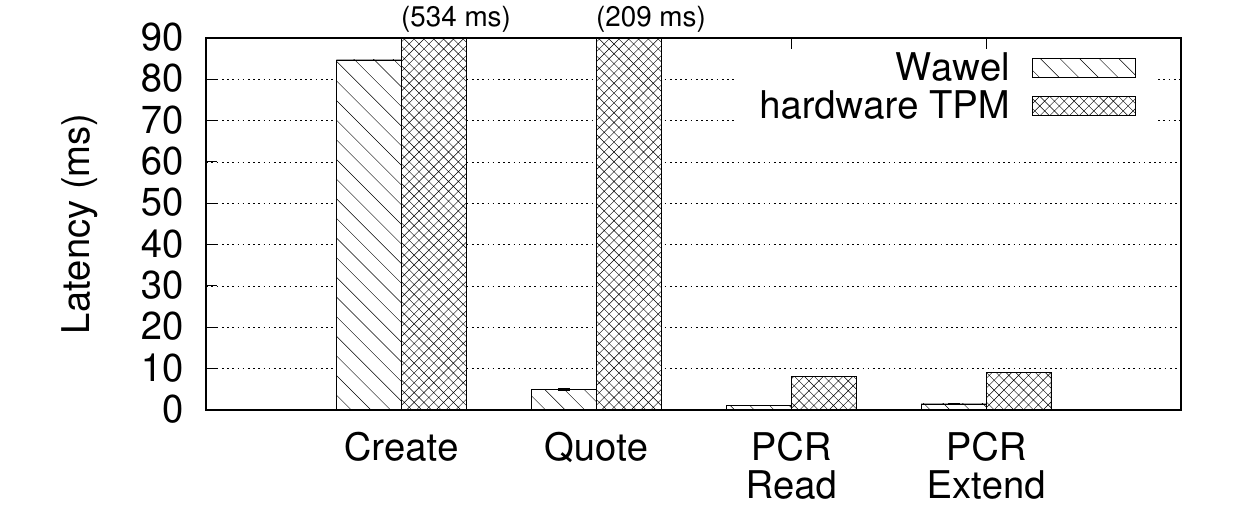}
    \caption{Latency of individual TPM operations executed on the \sys prototype.}
    \label{fig:eval:tpmoperations}
\end{figure}
\myparagraph{Does \sys has comparable latency to hardware TPM?}
To answer the question if \sys has an acceptable latency, we calculated the execution latency of TPM attestation-related commands in two variants: 1) a virtual machine equipped with \sys and 2) a virtual machine equipped with a hardware TPM. We measured the latency of the following TPM commands: \emph{Create}, \emph{Quote}, \emph{PCR Read}, \emph{PCR Extend}. The Create command is used to generate a new cryptographic key that can be used for quote signing. The Quote command retrieves from the TPM a signed attestation report over aggregated measurements. The PCR Read command returns aggregated measurements, while the PCR Extend cryptographically incorporates a measurement into the existing aggregated measurements. For each variant and TPM command, we ran the experiment ten times and calculated 10\%-trimmed mean as the resulting average of the TPM command latency. The hypervisor exposed the hardware TPM to the virtual machine via \texttt{passthrough}.

\autoref{fig:eval:tpmoperations} shows latency of TPM commands executed on \sys and a hardware TPM. The PCR Read and PCR Extend commands retrieve and store integrity measurements in TPM registers. The extend command is used by the booting firmware and \ac{IMA} to persist load-time and runtime integrity measurements. Thus, it is important to keep its latency as low as possible because it directly influences the boot time of the operating system and runtime applications. Latency of PCR Read and PCR Extend commands is lower for \sys, 1\,ms and 1.5\,ms, compared to hardware TPM, 8\,ms and 9\,ms, respectively. That means that despite the extra level of indirection and the network overhead, \sys is still faster than a locally attached hardware TPM. 

The Quote command triggers the generation and signing of a TPM quote containing the hash of selected PCRs. The hardware TPM requires 209\,ms to generate a quote, while the same operation takes less than 5\,ms on \sys. \sys outperforms the hardware TPM because the CryptoCard is optimized for performing low latency cryptographic operations, while hardware TPMs are bounded with the low price requirement. Quotes are signed with an attestation key created with the Create command. \sys is much more performant in executing the Create command because of the underlying hardware cryptographic accelerators and better source of high entropy. \sys required 84\,ms to execute the Create command while the hardware TPM 6.5$\times$ more, 534\,ms. 

\myparagraph{What is the overhead \sys on the boot time of a virtual machine?}
\begin{figure}[t!]
    \centering
    \includegraphics[width=0.5\textwidth]{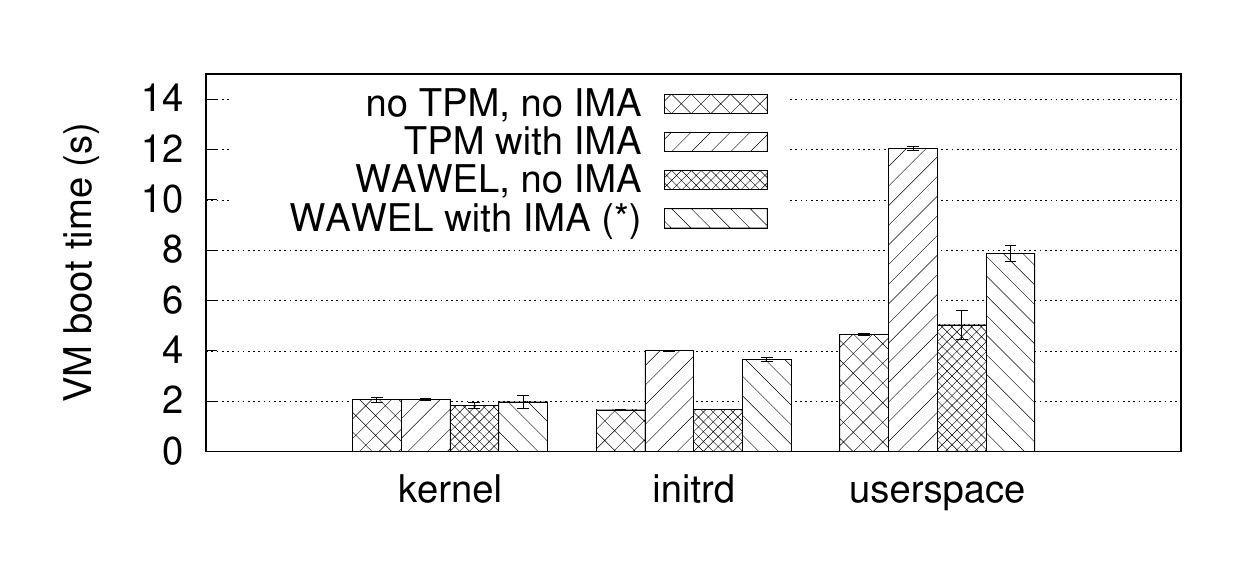}
    \caption{Impact of the \sys prototype on the virtual machine's boot time. (*) The experiment \emph{\sys with IMA} shows the time systemd took to boot the operating system. This time does not include the processing time of buffered TPM commands done by the proxy. }
    \label{fig:eval:boottime}
\end{figure}
We run the experiment to check how much the overhead introduced by the higher PCR Extend execution latency increases the virtual machine boot time. For that, we measured the boot time of a virtual machine in four variants: 
\begin{enumerate*}
    \item without a TPM and disabled IMA,
    \item with a hardware TPM and enabled IMA,
    \item with \sys but disabled IMA,
    \item with \sys and enabled IMA. 
\end{enumerate*}
We used the \emph{systemd-analyze} tool to collect the virtual machine boot time. For each variant, we ran the experiment ten times and then calculated a 10\% trimmed mean as the average time spent during the boot in the kernel, initrd, and userspace.

\autoref{fig:eval:boottime} shows the boot time of kernel, initrd, and userspace in each variant. The boot time of virtual machines with enabled IMA is longer because the kernel measures the integrity of files when loading them to memory and extends these measurements to the TPM. \sys achieves lower boot times of initrd and userspace because during the boot process it only buffers TPM commands without sending them to the hardware root of trust, \ie, the CryptoCard. This is because the CryptoCard is not accessible early in the boot due to the lack of network connectivity. Once the userspace is loaded, the proxy daemon sends buffered TPM commands to the CryptoCard. According to our measurements, all commands are processed by the CryptoCard after about 30\,sec, depending on the number of measurements. Please note that using the vsock interface could speed up this process because measurements could be processed without waiting for the network. However, it requires the support of the hypervisor that has to configure the vsock redirection to the shim.

\subsection{Scalability}

\begin{figure}[t!]
    \centering
    \includegraphics[width=0.5\textwidth]{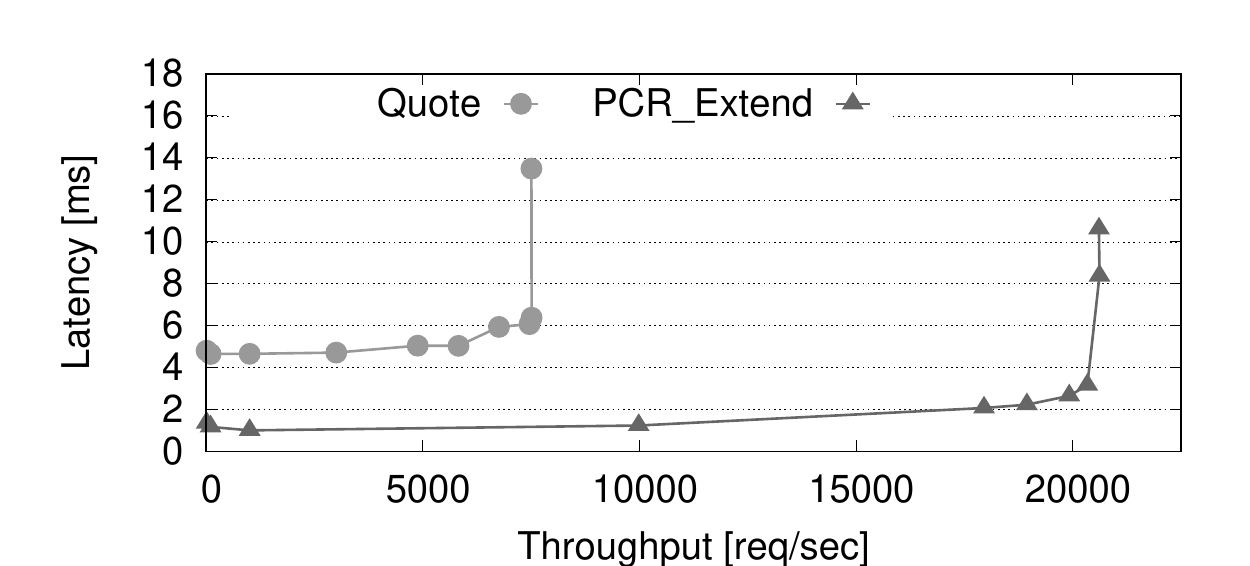}
    \caption{Throughput of the \sys prototype's shim executing PCR\_Extend and Quote commands. A single IBM CryptoCard acts as a backend.}
    \label{fig:eval:latput}
\end{figure}
\myparagraph{What is the \sys's throughput of aggregating integrity measurement and generating quotes?}
The previous experiments showed that \sys has lower latency than the locally attached TPM. However, a single TPM serves typically a single machine while in the \sys design, a limited number of CryptoCards virtualizes TPMs for much larger number of VEEs. To assess whether this approach is economically practical, we need to estimate how many VEEs can be supported by a single CryptoCard.

We ran an experiment in which we measured the throughput of two most frequently used TPM attestation-related commands: PCR Extend and Quote. In this experiment, we used the \emph{ghz} tool to generate desired gRPC throughput using eight connections to shim processes, which executed on the Machine B and had access to the CryptoCard over the PCIe interface. We were increasing the desired throughput until the latency increased over 10\,ms. We monitored the network to make sure the increased latency is not caused by the constrained network bandwidth.

\autoref{fig:eval:latput} shows the latency/throughput ratio of the PCR Extend and Quote commands. \sys achieved the throughput of 20k PCR extensions per second before reaching the saturation point. Considering that Linux IMA takes around 11\,ms to load and extend a single file of size lower than 100\,KiB~\cite{ozga2021triglav}, a single CryptoCard could handle around 222 parallel VEEs with enabled Linux IMA that constantly open, measure, and extend files measurement. In practice, Linux IMA extends the measurements of a file only the first time it reads it, remeasuring it only when it changes. The peak happens during the startup of the system when files are loaded to the memory for the first time. On Fedora 36, we collected around 2000 measurements using the \emph{ima\_tcb} policy that measures all files and executable owned by root. That means that a single CryptoCard could support up to 222 VEEs booting up at the same time for 22\,sec. 

\sys achieved a throughput of 7.5k quotes per second before reaching the saturation point. The existing integrity monitoring systems~\cite{intel_secl, ozga2022chors, ibm_tpm_acs} collect TPM quotes from monitored systems at regular configurable intervals at the resolution changing from seconds to hours. Depending on the interval configuration, a single CryptoCard could serve from 7.5k to 27k VEEs that are under constant integrity monitoring, assuming that quotes are collected every second or hour, respectively. For use cases where the attestation happens at specific point of time, for example during the establishment of an ssh~\cite{ozga2021triglav} or VPN~\cite{strongswan_org} connection, a single CryptoCard could handle up to 7.5k connections to/from a VEE per second.

\begin{figure}[t!]
    \centering
    \includegraphics[width=0.5\textwidth]{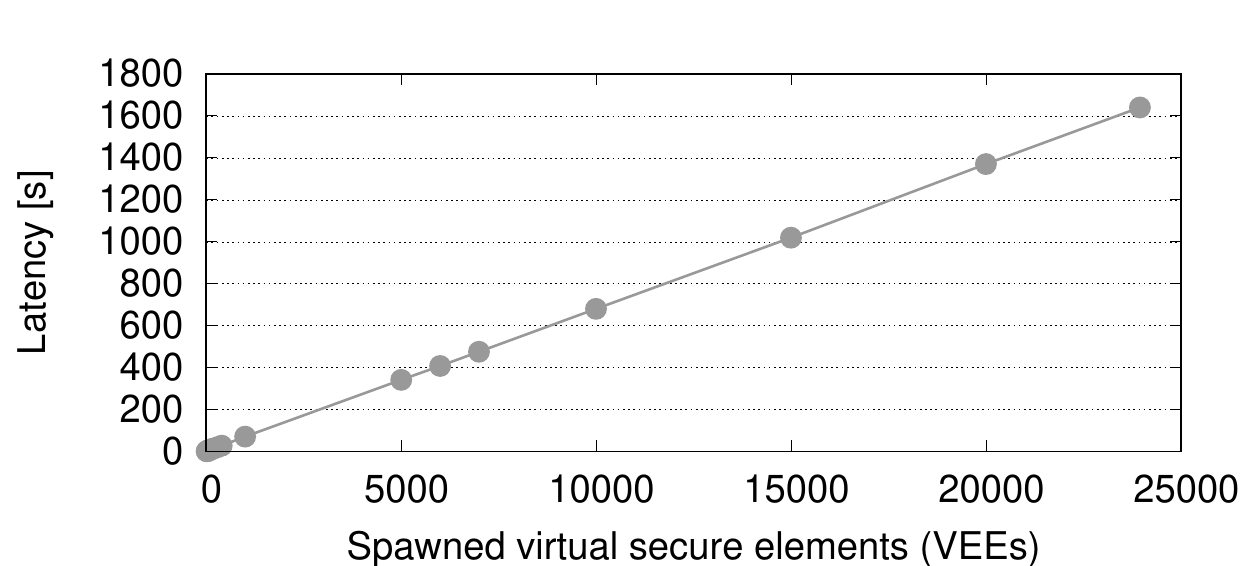}
    \caption{Scalability of spawning new VSE in the \sys prototype. A single VSE corresponds to a one VEE.}
    \label{fig:eval:scalability}
\end{figure}
\myparagraph{How effectively \sys handles increasing number of spawned VEEs?}
The cloud may spawn many VEEs simultaneously and each VEE might request its own VSE. In this experiment, we focused on verifying that \sys can effectively create new VSEs with the increasing number of clients. To check it, we ran an experiment in which we simulated \acp{VEE} that request the broker service for a new \acp{VSE}. The broker was then spawning a dedicated shim process that initialized the VSE state with the help of the CryptoCard. We simulated load using synchronous gRPC requests generated with the \emph{ghz} tool. The request was terminating successfully when the broker spawned a new shim process and the shim created a set of \acp{PCR} via the CryptoCard. 

\autoref{fig:eval:scalability} shows that the latency of spawning new \acp{VSE} grows linearly with the number of requested \ac{VSE} instances. We hit the limit of available computing resources when we spawned about 25k \acp{VSE}. This result indicates that \sys scales well for spawning new VSEs because \sys could handle more requests by adding more physical resources, like machines running shim instances and CryptoCards.

\begin{figure*}[btp]
    \centering
    % \captionsetup{skip=4pt}
    \includegraphics[width=\textwidth]{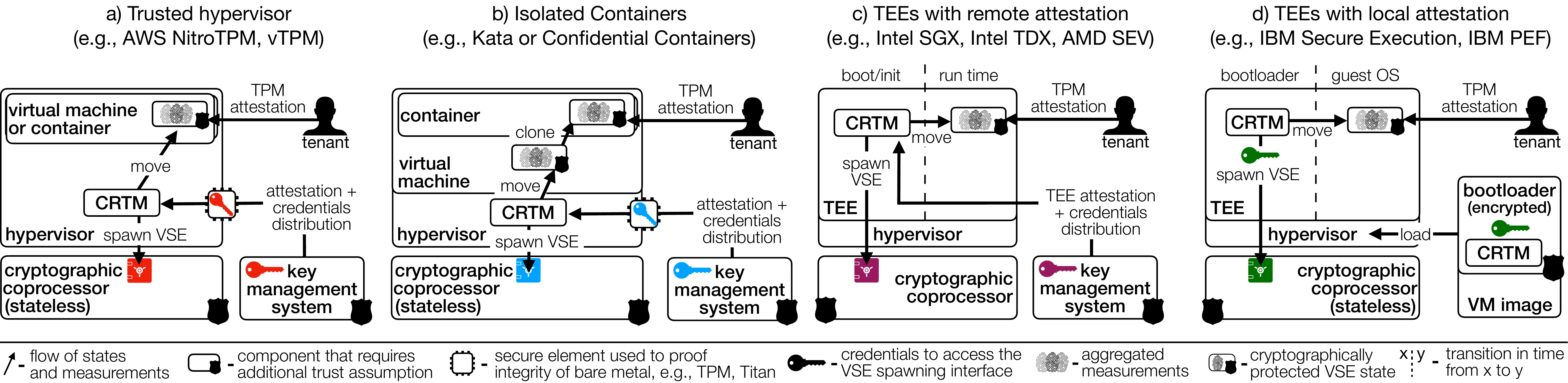}
    \caption{
        Integration of \sys with different virtualization and isolation technologies. \Acf{CRTM} is the trusted component that spawns new \acfp{VSE}. However, to spawn \acp{VSE} it authenticates itself with credentials, which it retrieves from key management systems or secure elements. The latter components ensure that they share credentials only with trusted, legitimate CRTMs.
    }
    \label{fig:relatedwork}
\end{figure*}
 
\section{Support for Heterogeneous VEEs and Related Work}
\label{sec:related_work}

\sys complements existing attestation and virtualization technologies, simplifying attestation in hybrid- and multi-clouds. The resulting threat models depend on the underneath technology. For example, enhancing \acf{TDX} \cite{tdxwhitepaper} with \sys provides a unified \ac{TPM}-based attestation protocol under the \ac{TDX} threat model with an untrusted hypervisor. Enhancing the legacy virtual machines with the TPM attestation supported by \sys requires placing trust in the hypervisor, like in the \ac{VTPM} design \cite{perez2006vtpm}. 

Tenants must differentiate among underlying virtualization technologies because these offer different security guarantees and trust assumptions. Thus, cryptographic coprocessors must certify the \acp{VEE} integrity with different attestation keys. This is possible by provisioning different \acp{CRTM} with access keys specific to the virtualization and isolation technology. Cryptographic coprocessors detect the virtualization technology when a \ac{CRTM} establishes a \ac{mTLS} connection to spawn a new \ac{VSE}. In \S\ref{sec:relatedwork:vtpm}, \S\ref{sec:relatedwork:tee}, and 
\S\ref{sec:relatedwork:dtpm}, we discuss \sys integration with various attestation and virtualization mechanisms used in modern clouds.

\subsection{Legacy Virtual Machines and Containers}
\label{sec:relatedwork:vtpm}

\autoref{fig:relatedwork}a) shows the integration of \sys with designs relying on trusted hypervisor whose integrity is measured by a hardware root of trust provided by a CPU-local secure element, like NitroTPM \cite{hamilton2021nitro}, Titan \cite{savagaonkar2017titan}, and \ac{TPM}. The secure element aggregates and certifies the host operating system's integrity in these designs. The host operating system exposes then software \acp{TPM} \cite{libtpms} to virtual machines to enable them with attestation primitives. The \ac{TCB} of this design includes the entire hypervisor and its operator. 

\sys complements this design by reducing the management effort of hosting software \acp{TPM} on the hypervisor. Each virtual machine manages its \ac{VSE} state, so there is no need for running emulated TPM processes on the hypervisor. The hypervisor, which is still part of the \ac{TCB}, implements the \ac{CRTM} that creates dedicated \acp{VSE} for each \ac{VEE}, such as legacy virtual machines and containers. Only the legitimate hypervisor can retrieve credentials enabling access to the \ac{VSE} creation interface. There are two ways for the hypervisor to retrieve credentials. In the first one, similarly to LUKS \cite{broz2018luks} or BitLocker, the hypervisor retrieves them from the secure element via the unseal mechanism (see, Section `PCRs for Authorization` in Chapter 12 in \cite{arthur2015tpm}). In the second one, it retrieves it from a key management system after proving its own identity and integrity using the hardware-specific secure element, like a hardware TPM.

\autoref{fig:relatedwork}b) shows a variant of this design that applies to hardware-isolated containers, such as Kata \cite{randazzo2019kata} or confidential containers \cite{confidentialcontainers}. The trusted component, like a hypervisor in the Kata container, implements the \ac{CRTM}. It spawns a dedicated \ac{VSE} instance for each virtual machine and extends it with the corresponding integrity measurements. Because all containers running inside a virtual machine share the same kernel, the virtual machine clones a \ac{VSE} for each container and extends it further with container-specific measurements. Like this, each container gets its own \ac{VSE} state derived from the parent's state representing its own and the underlying host integrity. The plurality of \acp{VSE} enables tracking of individual containers' runtime integrity via \ac{IMA} \cite{sailer2004ima}.

\subsection{Trusted Execution Environments}
\label{sec:relatedwork:tee}

\sys complements \acfp{TEE}, which are technologies that enable confidential computing by isolating \acp{VEE} from the hypervisor and operator. They offer attestation capabilities by measuring the load-time integrity of a \ac{VEE} and then certifying these measurements to the tenant. Unlike \sys, however, the TEE attestation mechanism is limited to vendor-specific hardware because hardware-specific keys sign the attestation certificate, and the hardware manufacturer controls these keys. \sys integration differs for TEEs supporting local or remote attestation. 

\autoref{fig:relatedwork}c) shows how \sys integrates with TEEs that support remote attestation, for example, Intel \ac{TDX} \cite{tdxwhitepaper}, Intel \ac{SGX} \cite{costan2016intel}, Sanctum \cite{costan2016sanctum}. The design integrating \sys with these technologies consists of three elements: the \ac{TEE} image containing the \ac{CRTM} and the tenant's application, the TEE-capable hardware, and the key management system containing credentials for accessing the \ac{VSE} creation interface. The key management system, like Palaemon \cite{gregor2020palaemon} or a key broker service in the confidential container design \cite{confidentialcontainers}, guards credentials and distributes them only to CRTMs after attesting to their integrity and identity using the TEE-specific remote attestation. The TEE-capable hardware certifies that the CRTM executes inside the TEE via the TEE-specific remote attestation, \eg, \cite{intel2018dcap, intel2016epid}. Consequently, the CRTM obtains credentials, which it first uses to spawn a new \ac{VSE}. Then, it extends \ac{VSE} with TEE- and application-specific measurements. It vanishes credentials before executing the tenant's application to prevent a malicious application from launching the reset attack discussed in \S\ref{sec:reset_attack}. The CRTM could be implemented as the init routine of a process-based TEE (\eg Palaemon~\cite{gregor2020palaemon}) or a part of the VM-based TEE's bootloader (\eg, confidential containers~\cite{confidentialcontainers}).

\autoref{fig:relatedwork}d) illustrates how \sys integrates with TEEs that support local attestation, for example, IBM Z Secure Execution \cite{borntraeger2020se} or IBM \ac{PEF} \cite{hunt2021confidential}. The local attestation differs from the remote attestation because the hardware verifies the TEE image during launch. In such TEEs, the tenant encrypts the \ac{VEE} image, or part of it, with a public key. The corresponding private key is known only to the hardware. The hardware TEE engine executes only \ac{VEE} encrypted with the public key corresponding to its private key. \sys integrates with this design by embedding the CRTM and credentials in the encrypted part of the \ac{VEE} image. The CRTM executes then as a first component of the \ac{VEE} inside the TEE. It creates a new \ac{VSE} by authenticating itself to the cryptographic coprocessor using the embedded credentials. Then, it extends the measurement of the \ac{VEE} image, removes the credentials, and executes the rest of the \ac{VEE}.

\subsection{Hardware TPM}
\label{sec:relatedwork:dtpm}
A hardware \ac{TPM} chip offers attestation primitives according to the TPM specification \cite{tcg2016tpm}. However, the TPM chip is a single device with limited storage that does not support virtualization. There were attempts to use it for attestation of virtual machines by emulating software TPMs \cite{perez2006vtpm} or running emulated TPMs inside of the TEE \cite{ozga2021triglav}. However, these technologies fit only specific use cases and do not apply for heterogeneous environments.  

Hardware secure elements, like hardware TPMs, Titan \cite{savagaonkar2017titan}, or Caliptra \cite{lagar2020caliptra}, apply to bare metal systems because they are available to boot firmware just after the processor is powered on. However, unlike \sys, they target only \acfp{VEE}. Like in the legacy systems relying on trusted hypervisors, see \autoref{fig:relatedwork}a) and \autoref{fig:relatedwork}b), \sys might reuse the boot integrity measurements of the bare metal. \sys can use the hardware TPM to attest to the bare metal integrity to unlock the CRTM access to credentials required for accessing the VSE creation interface. The \ac{CRTM} could then extend hardware TPM measurements into the \ac{VSE}, providing the continuity of the chain of trust to virtual environments.

\acresetall
\section{Conclusion}

This paper introduced \sys, a scalable attestation framework for \acp{VEE}, \eg, virtual machines and containers running on heterogeneous hardware and virtualization technologies. \sys reduces management effort because each \ac{VEE} provisions and maintains a \ac{VSE} with the help of a limited number of high-performant stateless cryptographic coprocessors. \sys scalability is achieved by offloading the cryptographic coprocessor state containing \ac{VEE} measurements securely, with the help of cryptography. 

Our prototype \sys implementation is compatible with the \ac{TPM} protocol and supports the \ac{IMA}, making it a drop-in replacement for existing virtual \ac{TPM} designs in the cloud. The evaluation results prove the scalability and practicality of the design for cloud deployements.

\bibliographystyle{acm}
\interlinepenalty=100
\bibliography{bibliography}

% \printbibliography

\end{document}